\documentclass[aps,pra,twocolumn,floatfix,floats,showpacs,superscriptaddress,raggedbottom]{revtex4-1}
\usepackage{graphicx,latexsym}
\usepackage{dcolumn}
\usepackage{amsmath}
\usepackage{braket}
\usepackage{amssymb,bm}
\usepackage{color}
\usepackage[normalem]{ulem}

\def\mb#1{\mbox{\boldmath$#1$}}

\def\sec#1{Sec.\ \ref{#1}}
\def\eq#1{Eq.\ (\ref{#1})}
\def\fig#1{Fig.\ \ref{#1}}

\usepackage{hyperref}
\hypersetup{
    pdfnewwindow=true,       
    colorlinks=true,         
    linkcolor=blue,          
    citecolor=blue,          
    filecolor=magenta,       
    urlcolor=black           
}

\begin{document}

\title{Optical switching of electron transport in a waveguide-QED system}

\author{Nzar Rauf Abdullah}
\email{nzar.r.abdullah@gmail.com}
\affiliation{Physics Department, Faculty of Science and Science Education, School of Science, 
             University of Sulaimani, Kurdistan Region, Iraq}
\affiliation{Science Institute, University of Iceland, 
             Dunhaga 3, IS-107 Reykjavik, Iceland}

\author{Chi-Shung Tang}
\affiliation{Department of Mechanical Engineering,
  National United University, 1, Lienda, Miaoli 36063, Taiwan}

\author{Andrei Manolescu}
\affiliation{Reykjavik University, School of Science and Engineering,
              Menntavegur 1, IS-101 Reykjavik, Iceland}

\author{Vidar Gudmundsson}
 \affiliation{Science Institute, University of Iceland,
        Dunhaga 3, IS-107 Reykjavik, Iceland}

%

\begin{abstract}
Electron switching in waveguides coupled to a photon cavity
is found to be strongly influenced by the photon energy and polarization.
Therefore, the charge dynamics
in the system is investigated in two different regimes, for off- and on-resonant photon fields. 
In the off-resonant photon field, the photon energy is smaller than the energy spacing between 
the first two lowest subbands of the waveguide system, the charge splits between 
the waveguides implementing a $\sqrt{\rm NOT}$-quantum logic gate action.
In the on-resonant photon field, the charge is totally switched from one waveguide to the other
due to the appearance of photon replica states of the first subband in the second subband region instigating 
a quantum-NOT transition.
In addition, the importance of the photon polarization to control the charge motion in the waveguide 
system is demonstrated.
The idea of charge switching in electronic circuits may serve to built quantum bits.
\end{abstract}



\maketitle

%
%

\section{Introduction}
\label{Sec:Introduction}

Quantum information processing is a rapidly growing field promising to harness the laws of quantum physics 
for the sake of improvements in computer technology~\cite{PhysRevLett.93.130502}.
One of the exciting aspects of the quantum information processing is the development of 
effective and fast computation strategies for data manipulation
with many possibilities~\cite{PhysRevA.70.052330}.
A number of different quantum systems are being explored to implement 
a quantum bit (qubit)~\cite{PhysRevLett.89.117901,PhysRevLett.84.5912,PhysRevA.74.012318}. 
Among these, double waveguides represented by a double quantum wire is a candidate to build
a qubit. A coupling element called coupling window is put between the waveguides 
to allow for inter-waveguide transport~\cite{PhysRevLett.84.5912}. In this system, the length of the coupling window 
can be tuned to form the proposed qubit and implement quantum logic gates~\cite{NANOTECHNOLOGYIEEE.6.5}.
Ionicioiu et al.\ have suggested the same scheme for quantum computation. 
In their scheme, a single electron is transported through 
a double quantum waveguide which states can be represented as qubit state $\ket{1}$ and $\ket{0}$~\cite{Ionicioiu.15.125}.
The manipulation of $\ket{0}$ to $\ket{1}$ or vice versa is performed by a quantum gate. 
The quantum NOT operation just negates the logical value: 0 becomes 1, and 
1 becomes 0. In addition, the superposition of 0 and 1 forms the $\sqrt{\rm{NOT}}$ 
quantum logic gate action. Several methods such as magnetic switching~\cite{ApplPhysLett.79.14}, 
electrically tunable~\cite{ApplPhysLett.81.22}, 
and split-gate method~\cite{Reno.19.276205} have been used as a quantum gate
to implement quantum logic actions in double waveguide system.

Recently, we investigated approaches to switch electron motion between two waveguides 
by a split-gate method, magnetic field, and photon cavity.
In addition, we reported the importance of the Coulomb interaction in the 
electron switching~\cite{nzar27.015301,Nzar116.23}.
In this work, we investigate the processes of electron switching in a coupled waveguide where a window 
coupling is placed between the waveguides.  
Our approach here is to use a single-photon mode to manipulate the electron motion between the two waveguides
in two regimes, for off- and resonant photon field.
The transient electron transport in the waveguide system is described using a generalized master equation~\cite{Vidar61.305,PhysRevB.81.155442}.
The switching processes implement the quantum logic gate actions in the double waveguide system representing the quantum bit.

This paper is organized as following: In \sec{Model and Theory} we define the model and theoretical methods. The results and conclusions are 
presented in \sec{Results} and \sec{Conclusions}, respectively.

\section{Model and Theory}
\label{Model and Theory}

\begin{figure}[htbq]
\centering
\includegraphics[width=0.45\textwidth]{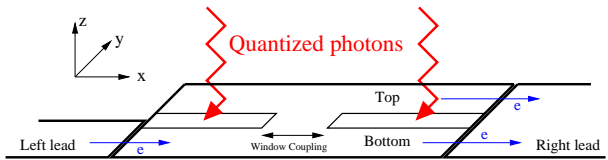}
\caption{(Color online) Schematic diagram shows two coupled waveguides connected to two leads.
                        The bottom waveguide is coupled to the left and the right leads 
                        while the top waveguide is only connected to the right lead.
                        Here, top and bottom refers to the y-direction.
                        The coupling window allows inter-waveguide transport.
                        The photon field is represented by red zigzag arrows and the blue arrows indicate 
                        the direction of electron motion in the system.}
\label{fig01}
\end{figure}

In this section, the model under investigation is introduced. 
We assume two symmetric waveguide system weakly coupled to two electron reservoirs or leads shown 
in \fig{fig01}. The bottom- and top-waveguide are so called control- and target-waveguide, respectively.
The waveguides are coupled via a coupling element called the coupling window that  
allows inter waveguide transport and electron interference between the waveguides. 
The control-waveguide is coupled to a lead from the left side and both waveguides are connected to a lead from right side.

The waveguide system is hard-wall confined in the $x$-direction and parabolically confined in
the $y$-direction. The waveguide potential can be described by 
\begin{eqnarray}
 U_{\rm WG}(r) = U_{0} \Big[ - e^{(-\alpha_0^2 x^2)} + e^{\left(-[\alpha_x^2 x^2 + \alpha_y^2 y^2]\right)} \Big].
\end{eqnarray}
Herein, $U_0$ is the strength of the confinement potential, and $\alpha_0$, $\alpha_x$ and $\alpha_y$ are constants.
The first term describes the potential barrier between the waveguides and the second is their coupling potential,
or coupling window.

The waveguide system is placed in a rectangular photon cavity with the cavity much larger than the waveguide system.
The cavity contains a single photon mode and the photons are linearly polarized in the cavity either parallel or 
perpendicular to the direction of the electron motion in the waveguide system.
The Hamiltonian of the double waveguide system coupled to a single photon mode 
in an external perpendicular magnetic field in the $z$-direction is~\cite{2040-8986-17-1-015201,PhysRevE.86.046701}
\begin{equation}\label{H_S}
 \hat{H}=\int d^2r\;\hat{\psi}^{\dagger}(\mathbf{r}) \left[\frac{\mathbf{\hat{P}}^2}{2m^{*}} + U_{\rm WG}(\mathbf{r}) \right] \hat{\psi}(\mathbf{r}) + 
              \hat{H}_{C} + \hat{H}_{\gamma},
\end{equation}
where $\hat{\psi}$ is the field operator, $m^*$ indicates the effective mass of an electron, and the canonical 
momentum operator is described as
\begin{equation}
 \mathbf{\hat{P}} = \left(\frac{\hbar}{i}\nabla+\frac{e}{c}\left[\mathbf{\hat{A}}(\mathbf{r}) +\mathbf{\hat{A}}_{\gamma}(\mathbf{r})\right] \right).
\end{equation}
Herein, $\mathbf{\hat{A}}(\mathbf{r})= -By \hat{\mb{x}}$ is the vector potential 
of the external constant magnetic field defined in the Landau gauge, and
$\hat{\mathbf{A}}_{\gamma}$ is the photonic vector potential of the cavity given by 
$\hat{\mathbf{A}}_{\gamma}(\mathbf{r})=A(\hat{a}+\hat{a}^{\dagger}) \mathbf{e}$, with
$A$ the amplitude of the photon field, $g_{\gamma}=eA a_w \Omega_w/c$ the electron-photon coupling constant,
$\mathbf{e} = \mathbf{e}_x$ for a longitudinally-polarized photon field ($\mathrm{TE}_{011}$),
or $\mathbf{e} = \mathbf{e}_y$ for 
transversely-polarized photon field ($\mathrm{TE}_{101}$),
and $\hat{a}$($\hat{a}^{\dagger}$) are annihilation(creation) operators of the photons, respectively.
The effective confinement frequency is $\Omega_w = \sqrt{\Omega^2_0 + \omega^2_c}$ with $\Omega_0$ the electron confinement frequency due to the lateral parabolic potential and $\omega_c$ the cyclotron frequency due to external magnetic field.
The confinement frequency defines a natural length scale $a_w=\sqrt{\hbar /(m^*\Omega_w)}$,
the effective magnetic length.
The second term of \eq{H_S} ($\hat{H}_{C}$)  
is the Coulomb interaction between the electrons in the waveguide system~\cite{Nzar.25.465302}.
The last term of \eq{H_S} represents the quantized
photon field $\hat{H}_{\gamma} = \hbar\omega_{\gamma} \hat{a}^{\dagger} \hat{a}$ 
with $\hbar\omega_{\gamma}$ the photon energy.
We investigate the electron transport properties in the waveguide-cavity system 
in the case of off- and on-resonant photon fields including both the para- and the diamagnetic 
electron-photon interactions without the rotating wave approximation~\cite{PhysRevE.86.046701}. 
The electron-electron and the electron-photon interactions are treated by exact diagonalization.

The waveguide system is coupled to external leads as indicated in Fig.\ \ref{fig01}
and described in earlier work~\cite{nzar27.015301,Nzar116.23}. 
The left and right leads are coupled simultaneously smoothly within 20 ps to the waveguide
system by the use of switching functions~\cite{Vidar61.305}.

We consider the leads to
be held at the same temperature $T$, and their overall coupling strength to the waveguides
is $g_\mathrm{LR}a^{3/2}_w$.

The continuous spectrum of the states in the leads which are treated as electron reservoirs, make the solution 
of the Liouville-von Neumann equation for the time evolution of the whole system impractical.
Instead, we use a formalism to project the time-evolution of the whole system onto the open
system consisting of the waveguides and the cavity, transforming the Liouville-von Neumann equation
for the full density operator to a generalized non-Markovian master equation for the reduced density 
operator (RDO)~\cite{Nakajima20.948,Zwanzing.33.1338}. This operation introduces complicated dissipation 
terms into the equation of motion describing the loss or gain of electrons and energy from the 
leads. The master equation describes the time-evolution of the RDO 
for the open waveguide system under the influences of the leads.
The non-Markovian approach allows us to study the transient behavior of the system in the 
weak coupling of the leads to the waveguide system~\cite{Vidar61.305}.

\section{Results}
\label{Results}

We consider two parallel waveguides made of a GaAs semiconductor material 
with electron effective mass $m^* = 0.067m_e$ and the relative dielectric constant $k = 12.4$.
The waveguide system with length $L_x = 300$~nm is weakly coupled to two electron reservoirs.
The transverse confinement energy of the electrons in the waveguide system is equal to that of the 
leads $\hbar\Omega_0 = \hbar\Omega_l = 1.0$~meV, where $l$ stands for the left (L), or the right (R) lead. 
The temperature of the leads is assumed to be $T_l = 0.5$~K.
The parameters that specify the potential barrier and the coupling window between the waveguide are 
$U_0 = 18.0$~meV, and $\alpha_0 = \alpha_y = 0.03$~nm$^{-1}$. The coupling window length is defined by
$L_{\rm CL} = 2/\alpha_x$, and $g_\mathrm{LR}a^{3/2}_w= 0.5$ meV. 
In addition, we assume the cavity initially contains one photon.

\subsection{Off-resonance photon field}\label{sec:2}

In this section, we consider the photon energy smaller than the energy 
spacing between the first and the second subband of the 
waveguides. The system under this condition is in the so called off-resonance regime.

\begin{figure}[htbq]
\centering
\includegraphics[width=0.35\textwidth]{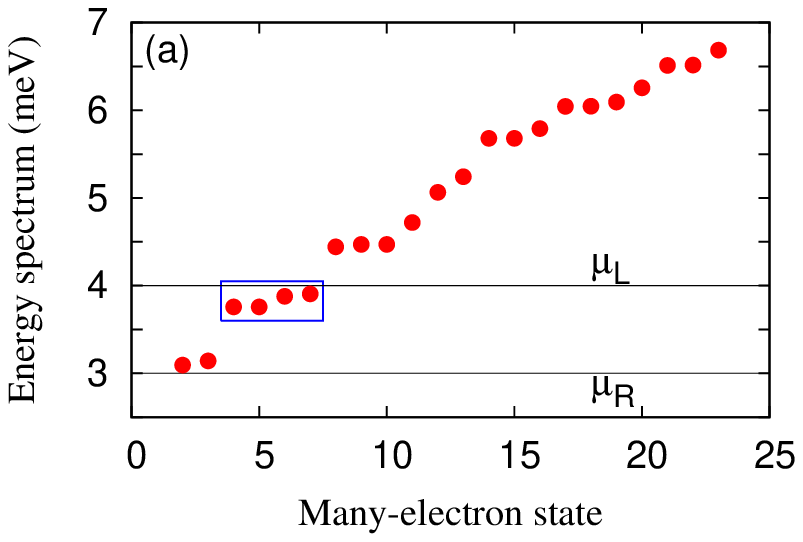}
\includegraphics[width=0.35\textwidth]{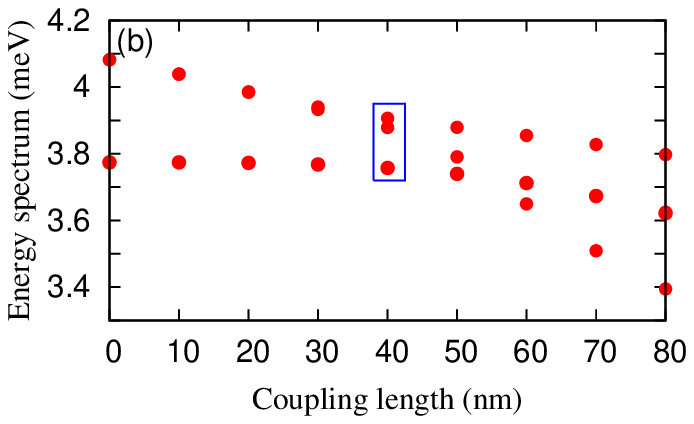}
\caption{(Color online) (a) Energy spectrum of the waveguide system versus many-electron state is plotted at $L_{\rm CL} = 40$~nm.
                            The black lines display the chemical potential of the left ($\mu_L$) and the right ($\mu_R$) lead.
                            The states in the blue rectangle are active states in the transport.
                        (b) Energy spectrum as a function of the coupling length in the case of no electron-photon coupling. 
                            The crossover in the energy spectrum shown in the blue rectangles indicates a proper
                            coupling between the waveguides. The magnetic field is $B = 0.1$~T, $\hbar\Omega_0 = 1.0$~meV.
                            The chemical potentials are $\mu_L = 3.0$~meV and $\mu_L = 4.0$~meV implying $\Delta \mu = 1.0$~meV.}
\label{fig02}
\end{figure}

First, we show the results of the waveguide system without the photon cavity.
Figure \ref{fig02} shows the energy spectrum of the waveguide system with no electron-photon coupling.
In \fig{fig02}(a) the many-electron (ME) energy versus the ME state $\rvert \mu)$ is plotted. The black lines 
indicate the chemical potential of the left ($\mu_L$) and the right ($\mu_R$) lead.
It can be clearly seen that the first subband of the waveguide system is located in the bias 
window $\Delta \mu = \mu_L - \mu_R$ including 
six one-electron states. We can confirm that only
four of them shown in the blue rectangle are active in the electron transport. The two lowest states, 
the ground state and the first-excited state, only weakly participate in the electron transport because of their electron
localization property in the coupling window region.
The four states are: second-, third-, fourth-, and fifth-excited states. The second subband contains six more states in the 
energy range $4.5\text{-}5.5$~meV.

The coupling window between the waveguides can be varied to find a suitable coupling between the waveguides. Practically, 
the coupling window can be formed using 'finger' gates or a saddle potential~\cite{NANOTECHNOLOGYIEEE.6.5}.
In \fig{fig02}(b) the energy spectrum of the four active states shown in~\fig{fig02}(a) 
versus the length of the coupling window is shown.
We notice a crossing/anti-crossing (blue rectangle) in the energy spectrum occurring at length $L_{\rm CL} = 40$~nm 
for the coupling window. 
The crossing point in the energy spectrum indicates a strong coupling or interference between the waveguides~\cite{PhysRevB.76.195301}. 

\begin{figure}[htbq]
\centering
\includegraphics[width=0.3\textwidth]{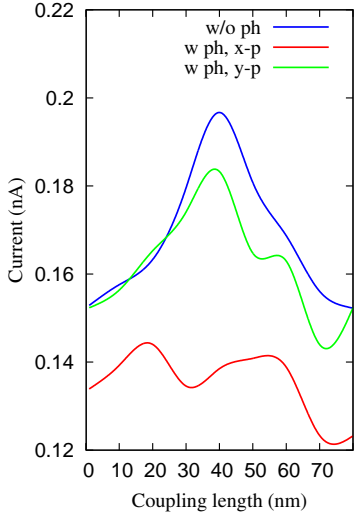}
\caption{(Color online) The net charge current versus coupling length of the waveguide system without (w/o) photon 
                        (ph) (blue lines)
                        and with photon (w ph) cavity in the case of $x$-polarization (red lines) and $y$-polarization (green lines) 
                        of the photons in the case of the off-resonance 
                        photon field. The magnetic field is $B = 0.1$~T, $\hbar\Omega_0 = 1.0$~meV.
                        The chemical potentials are $\mu_L = 3.0$~meV and $\mu_L = 4.0$~meV implying $\Delta \mu = 1.0$~meV.}
\label{fig03}
\end{figure}

\begin{figure}[htbq]
\centering
\includegraphics[width=0.45\textwidth]{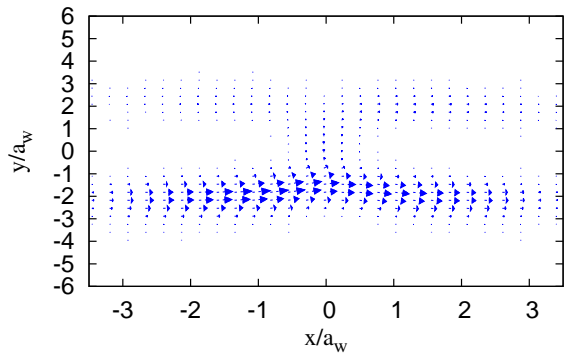}
\caption{(Color online) Charge current density in the waveguide system 
                        for the current peak at $t = 200$~ps and $L_{\rm CL} = 40$~nm 
                        shown in \fig{fig03} (blue lines).
                        The electron-photon interaction is neglected here.
                        The magnetic field is $B = 0.001$~T and the 
                        effective magnetic length is $a_w = 33.72$~nm.}
\label{fig04}
\end{figure}

To illustrate the physical properties of electron transport in the crossing region, we display
the net charge current and current density in the late transient regime at time $t = 200$~ps.
\fig{fig03} demonstrates the net charge current versus the coupling length for the waveguide system 
without (w/o) photon (ph) (blue lines)
and with photon (w ph) cavity in the case of an $x$-polarized (red lines) and a $y$-polarized (green lines) 
photon field.
The current is maximum at $L_{\rm CL} = 40$~nm in the energy spectrum for the case of no 
electron-photon coupling (blue curve) corresponding to the crossing in the energy spectrum. 
The maximized current at $40$~nm can be explained by observing charge current density.  
In a previous work, we have shown the dynamic motion of charge through the waveguide system in the absence of the 
Coulomb interaction~\cite{nzar27.015301} in which the incoming charge from the input bottom waveguide 
is equally split between the two outputs of the waveguides at $L_{\rm CL} = 40$~nm. 
The splitting process of the charge density occurs due to a contribution of 
two electron states to the transport. But the Coulomb interaction breaks the splitting process 
by lifting the two electron states above the group of active states.
Therefore, the charge density flows from the left lead to the right lead through the bottom 
waveguide without inter-waveguide backward or forward scattering as is shown in \fig{fig04}. 
The current is thus maximized and a current peak is seen for $L_{\rm CL} = 40$~nm. 

\begin{figure}[htbq]
\centering
\includegraphics[width=0.45\textwidth,angle=0,bb=50 110 300 220]{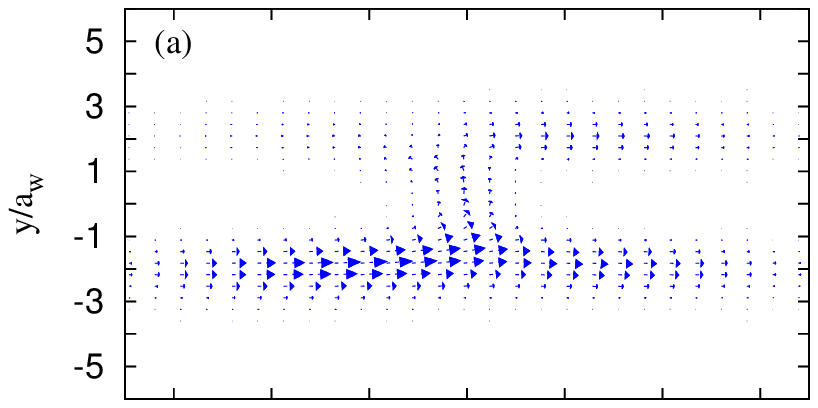}
\includegraphics[width=0.45\textwidth,angle=0,bb=50 60 300 220]{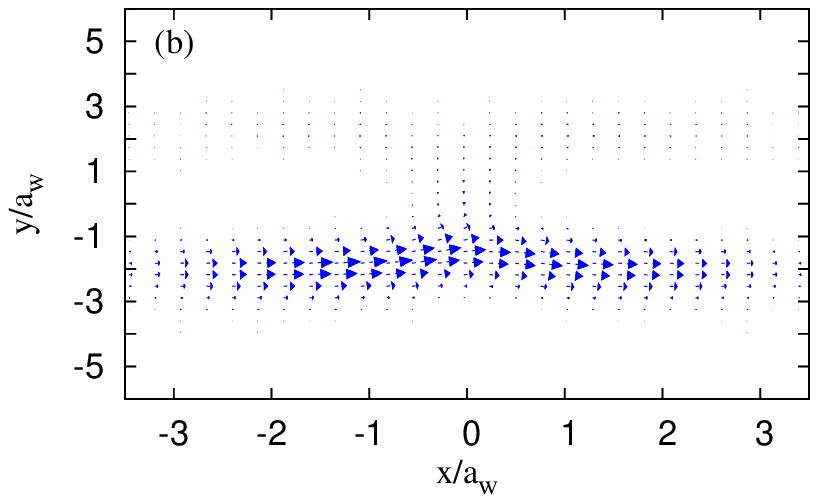}
\caption{(Color online) Charge current density in the waveguide system at $t = 200$~ps and $L_{\rm CL} = 40$~nm 
                        for the system coupled to the photon cavity with $x$-polarization (a) and $y$-polarization (b) 
                        of the photon field corresponding to the current peak labeled as red line and green 
                        line shown in \fig{fig03}, respectively.
                        The photon energy is $\hbar\omega_{\gamma} = 0.5$~meV and $g_{\gamma} = 0.1$~meV.
                        The magnetic field is $B = 0.001$~T and the effective magnetic length is $a_w = 33.72$~nm.}
\label{fig05}
\end{figure}

Now, we consider the waveguide to be embedded in the photon cavity. The photons are linearly polarized in 
either $x$ or $y$-direction. 
As we mentioned above, the photon energy is smaller than the energy spacing between the first and the second 
subband of the waveguide system.
In the presence of the photon cavity, photon replica states are formed and they actively participate in the electron transport~\cite{Nzar116.23}. 
For the case of an off-resonant photon field we choose the energy $\hbar\Omega_{\gamma} = 0.5$~meV,
which is smaller than the electron confinement energy of the waveguide system in the 
$y$-direction ($\hbar\Omega_0 = 1.0$~meV). 
In this case, the one-photon replicas of the four active states mentioned above are formed between the first 
and second subbands in the energy range $[4.0 \text{-} 4.5]$~meV at $L_{\rm CL} = 40$~nm.
The participation of the photon replica states modifies the charge motion in the system. 
The dynamic motion of the charge at $L_{\rm CL} = 40$~nm demonstrated in \fig{fig05}(a) indicates that it
splits between the top and the bottom waveguides. The charge splitting here implements a quantum logic 
gate called a $\sqrt{\rm NOT}$-gate.
This is analogue to the superposition of 
the ground state and exited state in a simple two level system in which 
the electron transmitted through both the $\bra{0}$ and the $\bra{1}$ states~\cite{Nielsen2010}.
We notice that the current is decreased in the $x$-polarized case of the photon field as is shown 
in \fig{fig03}(a) (red lines).

Figure \ref{fig05}(b) displays the charge current density for the case of a $y$-polarized photon field.
The charge is transported through the bottom waveguide without much inter-waveguide transport.
Consequently, the current of the peak is slightly decreased as is displayed in \fig{fig03}(a) (green lines).

The differing 'conductance' of the system with respect to the photon polarization 
reflects the geometric anisotropy of the waveguide system.

\subsection{Resonant photon field}

\begin{figure}[htbq]
\centering
\includegraphics[width=0.45\textwidth,angle=0,bb=50 110 300 220]{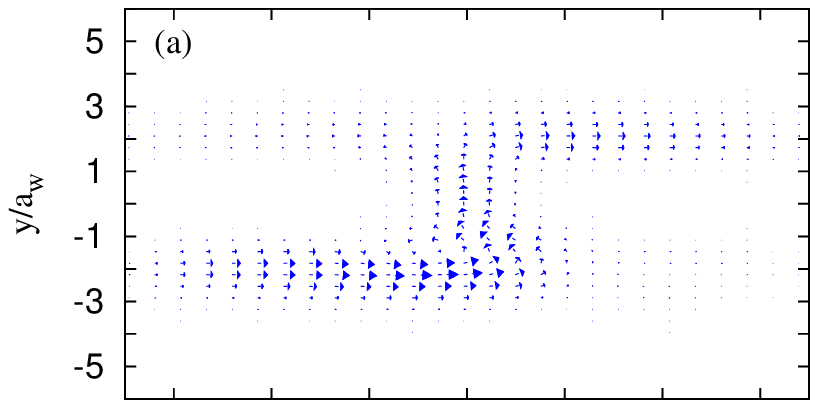}
\includegraphics[width=0.45\textwidth,angle=0,bb=50 60 300 220]{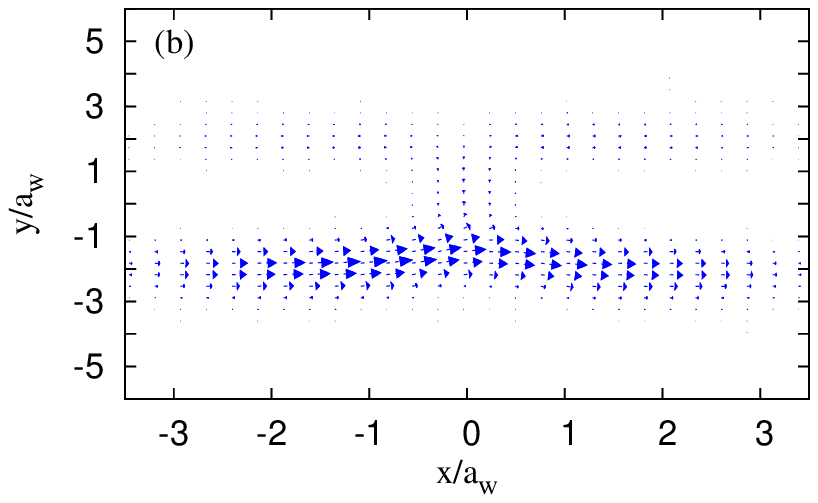}
\caption{(Color online) Charge current density in the waveguide system at $t = 200$~ps and $L_{\rm CL} = 40$~nm 
                        for the system coupled to the photon cavity with $x$-polarization (a) and $y$-polarization (b) 
                        of the photon field. The photon energy is $\hbar\omega_{\gamma} = 0.7$~meV and $g_{\gamma} = 0.1$~meV.
                        The magnetic field is $B = 0.001$~T and the effective magnetic length is $a_w = 33.72$~nm.}
\label{fig06}
\end{figure}

In this section, we increase the photon energy to  $\hbar\omega_{\gamma} = 0.7$~meV. 
However, the photon energy is still a bit smaller than the electron confinement energy 
of the waveguide system $\hbar \Omega_0 = 1.0$~meV, but we can obtain a total charge switching 
between the waveguides.
The one photon replicas of the four active states mentioned in the previous section are now formed in the second subband 
when the photon energy is $0.7$~meV. The second subband of the waveguide system becomes active in the electron transport. 
Consequently, the transmission of charge from the bottom guide input to 
the top guide output at $L_{\rm CL} = 40$~nm is obtained in the $x$-polarized of the photon field
as is shown in \fig{fig06}(a).
The switching of the charge transport implements a NOT-operation quantum logic gate.
But, for a $y$-polarized photon field presented in \fig{fig06}(b), 
the charge current is almost unchanged.

\section{Conclusions}
\label{Conclusions}

The results achieved in this study lead to the conclusion that, by applying an optical source such as photons
in a cavity to coupled waveguides, it is possible to stimulate electron switching between the two waveguides. 
A coherent propagation of an electron along a waveguide and a subsequent charge switching has been achieved.
We initialize the double waveguide system by injecting a single electron in one of the waveguide, this is done 
by coupling one of the waveguides to a lead on the left side. Calculation of the final state is performed by coupling 
both waveguides to a lead on the right side. The length of the inter-waveguide coupling window 
is tuned leading to a crossing point in the energy spectrum.
The crossing indicates a strong coupling between the waveguides.
The net charge current is thus at a maximum in the crossing region.
By applying linearly polarized cavity photons to the electronic system in parallel quantum waveguides, 
we demonstrate the implementation of two types of quantum gates in the energy crossing region
reflecting strong coupling of the waveguides.
For an off-resonant $x$-polarized photon field, the splitting of the charge transfer
has been found to be caused by a formation 
of photon replica states leading to a $\sqrt{\rm NOT}$-gate operation.
For the on-resonant and $x$-polarized photon field, the charge transfer switches from one waveguide to the other.
The motion of charge implements a NOT-operation quantum logic gate.
We like to underline that the charge motion is not influenced by the photon field 
in the case of off- or on-resonance for the case of y-polarization due to the geometric anisotropy of the waveguide system.
With this study we point out the possibility to use a waveguide-photon cavity system
to implement the fundamental qubit operations needed for quantum information processing.
Importantly, our time-dependent non-Markovian calculations point to the possibility to achieve the 
qubit switching in the late transient regime of the system.

\begin{acknowledgments}

Financial support is acknowledged from the Icelandic Research and Instruments Funds, 
and the Research Fund of the University of Iceland. The calculations were carried out on 
the Nordic High Performance Computer Center in Iceland. We acknowledge the Nordic network
NANOCONTROL, project No.: P-13053, University of Sulaimani, and 
the Ministry of Science and Technology, Taiwan 
through Contract No. MOST 103-2112-M-239-001-MY3.

\end{acknowledgments}


\begin{thebibliography}{21}%
\makeatletter
\providecommand \@ifxundefined [1]{%
 \@ifx{#1\undefined}
}%
\providecommand \@ifnum [1]{%
 \ifnum #1\expandafter \@firstoftwo
 \else \expandafter \@secondoftwo
 \fi
}%
\providecommand \@ifx [1]{%
 \ifx #1\expandafter \@firstoftwo
 \else \expandafter \@secondoftwo
 \fi
}%
\providecommand \natexlab [1]{#1}%
\providecommand \enquote  [1]{``#1''}%
\providecommand \bibnamefont  [1]{#1}%
\providecommand \bibfnamefont [1]{#1}%
\providecommand \citenamefont [1]{#1}%
\providecommand \href@noop [0]{\@secondoftwo}%
\providecommand \href [0]{\begingroup \@sanitize@url \@href}%
\providecommand \@href[1]{\@@startlink{#1}\@@href}%
\providecommand \@@href[1]{\endgroup#1\@@endlink}%
\providecommand \@sanitize@url [0]{\catcode `\\12\catcode `\$12\catcode
  `\&12\catcode `\#12\catcode `\^12\catcode `\_12\catcode `\%12\relax}%
\providecommand \@@startlink[1]{}%
\providecommand \@@endlink[0]{}%
\providecommand \url  [0]{\begingroup\@sanitize@url \@url }%
\providecommand \@url [1]{\endgroup\@href {#1}{\urlprefix }}%
\providecommand \urlprefix  [0]{URL }%
\providecommand \Eprint [0]{\href }%
\providecommand \doibase [0]{http://dx.doi.org/}%
\providecommand \selectlanguage [0]{\@gobble}%
\providecommand \bibinfo  [0]{\@secondoftwo}%
\providecommand \bibfield  [0]{\@secondoftwo}%
\providecommand \translation [1]{[#1]}%
\providecommand \BibitemOpen [0]{}%
\providecommand \bibitemStop [0]{}%
\providecommand \bibitemNoStop [0]{.\EOS\space}%
\providecommand \EOS [0]{\spacefactor3000\relax}%
\providecommand \BibitemShut  [1]{\csname bibitem#1\endcsname}%
\let\auto@bib@innerbib\@empty
\bibitem [{\citenamefont {M\"{o}tt\"{o}nen}\ \emph {et~al.}(2004)\citenamefont
  {M\"{o}tt\"{o}nen}, \citenamefont {Vartiainen}, \citenamefont {Bergholm},\
  and\ \citenamefont {Salomaa}}]{PhysRevLett.93.130502}%
  \BibitemOpen
  \bibfield  {author} {\bibinfo {author} {\bibfnamefont {M.}~\bibnamefont
  {M\"{o}tt\"{o}nen}}, \bibinfo {author} {\bibfnamefont {J.~J.}\ \bibnamefont
  {Vartiainen}}, \bibinfo {author} {\bibfnamefont {V.}~\bibnamefont
  {Bergholm}}, \ and\ \bibinfo {author} {\bibfnamefont {M.~M.}\ \bibnamefont
  {Salomaa}},\ }\href {\doibase 10.1103/PhysRevLett.93.130502} {\bibfield
  {journal} {\bibinfo  {journal} {Phys. Rev. Lett.}\ }\textbf {\bibinfo
  {volume} {93}},\ \bibinfo {pages} {130502} (\bibinfo {year}
  {2004})}\BibitemShut {NoStop}%
\bibitem [{\citenamefont {Snyder}\ and\ \citenamefont
  {Reichl}(2004)}]{PhysRevA.70.052330}%
  \BibitemOpen
  \bibfield  {author} {\bibinfo {author} {\bibfnamefont {M.~G.}\ \bibnamefont
  {Snyder}}\ and\ \bibinfo {author} {\bibfnamefont {L.~E.}\ \bibnamefont
  {Reichl}},\ }\href {\doibase 10.1103/PhysRevA.70.052330} {\bibfield
  {journal} {\bibinfo  {journal} {Phys. Rev. A}\ }\textbf {\bibinfo {volume}
  {70}},\ \bibinfo {pages} {052330} (\bibinfo {year} {2004})}\BibitemShut
  {NoStop}%
\bibitem [{\citenamefont {Martinis}\ \emph {et~al.}(2002)\citenamefont
  {Martinis}, \citenamefont {Nam}, \citenamefont {Aumentado},\ and\
  \citenamefont {Urbina}}]{PhysRevLett.89.117901}%
  \BibitemOpen
  \bibfield  {author} {\bibinfo {author} {\bibfnamefont {J.~M.}\ \bibnamefont
  {Martinis}}, \bibinfo {author} {\bibfnamefont {S.}~\bibnamefont {Nam}},
  \bibinfo {author} {\bibfnamefont {J.}~\bibnamefont {Aumentado}}, \ and\
  \bibinfo {author} {\bibfnamefont {C.}~\bibnamefont {Urbina}},\ }\href
  {\doibase 10.1103/PhysRevLett.89.117901} {\bibfield  {journal} {\bibinfo
  {journal} {Phys. Rev. Lett.}\ }\textbf {\bibinfo {volume} {89}},\ \bibinfo
  {pages} {117901} (\bibinfo {year} {2002})}\BibitemShut {NoStop}%
\bibitem [{\citenamefont {Bertoni}\ \emph {et~al.}(2000)\citenamefont
  {Bertoni}, \citenamefont {Bordone}, \citenamefont {Brunetti}, \citenamefont
  {Jacoboni},\ and\ \citenamefont {Reggiani}}]{PhysRevLett.84.5912}%
  \BibitemOpen
  \bibfield  {author} {\bibinfo {author} {\bibfnamefont {A.}~\bibnamefont
  {Bertoni}}, \bibinfo {author} {\bibfnamefont {P.}~\bibnamefont {Bordone}},
  \bibinfo {author} {\bibfnamefont {R.}~\bibnamefont {Brunetti}}, \bibinfo
  {author} {\bibfnamefont {C.}~\bibnamefont {Jacoboni}}, \ and\ \bibinfo
  {author} {\bibfnamefont {S.}~\bibnamefont {Reggiani}},\ }\href {\doibase
  10.1103/PhysRevLett.84.5912} {\bibfield  {journal} {\bibinfo  {journal}
  {Phys. Rev. Lett.}\ }\textbf {\bibinfo {volume} {84}},\ \bibinfo {pages}
  {5912} (\bibinfo {year} {2000})}\BibitemShut {NoStop}%
\bibitem [{\citenamefont {Reichl}\ and\ \citenamefont
  {Snyder}(2006)}]{PhysRevA.74.012318}%
  \BibitemOpen
  \bibfield  {author} {\bibinfo {author} {\bibfnamefont {L.~E.}\ \bibnamefont
  {Reichl}}\ and\ \bibinfo {author} {\bibfnamefont {M.~G.}\ \bibnamefont
  {Snyder}},\ }\href {\doibase 10.1103/PhysRevA.74.012318} {\bibfield
  {journal} {\bibinfo  {journal} {Phys. Rev. A}\ }\textbf {\bibinfo {volume}
  {74}},\ \bibinfo {pages} {012318} (\bibinfo {year} {2006})}\BibitemShut
  {NoStop}%
\bibitem [{\citenamefont {Ramamoorthy}\ \emph {et~al.}(2006)\citenamefont
  {Ramamoorthy}, \citenamefont {Akis},\ and\ \citenamefont
  {Bird}}]{NANOTECHNOLOGYIEEE.6.5}%
  \BibitemOpen
  \bibfield  {author} {\bibinfo {author} {\bibfnamefont {A.}~\bibnamefont
  {Ramamoorthy}}, \bibinfo {author} {\bibfnamefont {R.}~\bibnamefont {Akis}}, \
  and\ \bibinfo {author} {\bibfnamefont {J.~P.}\ \bibnamefont {Bird}},\
  }\href@noop {} {\bibfield  {journal} {\bibinfo  {journal} {IEEE Transaction
  on nanotechnology}\ }\textbf {\bibinfo {volume} {5}},\ \bibinfo {pages} {712}
  (\bibinfo {year} {2006})}\BibitemShut {NoStop}%
\bibitem [{\citenamefont {Ionicioiu}\ \emph {et~al.}(2001)\citenamefont
  {Ionicioiu}, \citenamefont {Amaratunga},\ and\ \citenamefont
  {Udrea}}]{Ionicioiu.15.125}%
  \BibitemOpen
  \bibfield  {author} {\bibinfo {author} {\bibfnamefont {R.}~\bibnamefont
  {Ionicioiu}}, \bibinfo {author} {\bibfnamefont {G.}~\bibnamefont
  {Amaratunga}}, \ and\ \bibinfo {author} {\bibfnamefont {F.}~\bibnamefont
  {Udrea}},\ }\href@noop {} {\bibfield  {journal} {\bibinfo  {journal} {Int. J.
  of Mod. Phys. B}\ }\textbf {\bibinfo {volume} {15}},\ \bibinfo {pages} {125}
  (\bibinfo {year} {2001})}\BibitemShut {NoStop}%
\bibitem [{\citenamefont {Harris}\ \emph {et~al.}(2001)\citenamefont {Harris},
  \citenamefont {Akis},\ and\ \citenamefont {Ferry}}]{ApplPhysLett.79.14}%
  \BibitemOpen
  \bibfield  {author} {\bibinfo {author} {\bibfnamefont {J.}~\bibnamefont
  {Harris}}, \bibinfo {author} {\bibfnamefont {R.}~\bibnamefont {Akis}}, \ and\
  \bibinfo {author} {\bibfnamefont {D.~K.}\ \bibnamefont {Ferry}},\ }\href@noop
  {} {\bibfield  {journal} {\bibinfo  {journal} {Appl. Phys. Lett.}\ }\textbf
  {\bibinfo {volume} {79}},\ \bibinfo {pages} {2214} (\bibinfo {year}
  {2001})}\BibitemShut {NoStop}%
\bibitem [{\citenamefont {Gilbert}\ \emph {et~al.}(2002)\citenamefont
  {Gilbert}, \citenamefont {Akis},\ and\ \citenamefont
  {Ferry}}]{ApplPhysLett.81.22}%
  \BibitemOpen
  \bibfield  {author} {\bibinfo {author} {\bibfnamefont {M.~J.}\ \bibnamefont
  {Gilbert}}, \bibinfo {author} {\bibfnamefont {R.}~\bibnamefont {Akis}}, \
  and\ \bibinfo {author} {\bibfnamefont {D.~K.}\ \bibnamefont {Ferry}},\
  }\href@noop {} {\bibfield  {journal} {\bibinfo  {journal} {Appl. Phys.
  Lett.}\ }\textbf {\bibinfo {volume} {81}},\ \bibinfo {pages} {4284} (\bibinfo
  {year} {2002})}\BibitemShut {NoStop}%
\bibitem [{\citenamefont {Ramamoorthy}\ \emph {et~al.}(2007)\citenamefont
  {Ramamoorthy}, \citenamefont {Bird},\ and\ \citenamefont
  {Reno}}]{Reno.19.276205}%
  \BibitemOpen
  \bibfield  {author} {\bibinfo {author} {\bibfnamefont {A.}~\bibnamefont
  {Ramamoorthy}}, \bibinfo {author} {\bibfnamefont {J.~P.}\ \bibnamefont
  {Bird}}, \ and\ \bibinfo {author} {\bibfnamefont {J.~L.}\ \bibnamefont
  {Reno}},\ }\href@noop {} {\bibfield  {journal} {\bibinfo  {journal} {Journal
  of Physics:Condensed Matter}\ }\textbf {\bibinfo {volume} {19}},\ \bibinfo
  {pages} {276205} (\bibinfo {year} {2007})}\BibitemShut {NoStop}%
\bibitem [{\citenamefont {Abdullah}\ \emph {et~al.}(2015)\citenamefont
  {Abdullah}, \citenamefont {Tang}, \citenamefont {Manolescu},\ and\
  \citenamefont {Gudmundsson}}]{nzar27.015301}%
  \BibitemOpen
  \bibfield  {author} {\bibinfo {author} {\bibfnamefont {N.~R.}\ \bibnamefont
  {Abdullah}}, \bibinfo {author} {\bibfnamefont {C.~S.}\ \bibnamefont {Tang}},
  \bibinfo {author} {\bibfnamefont {A.}~\bibnamefont {Manolescu}}, \ and\
  \bibinfo {author} {\bibfnamefont {V.}~\bibnamefont {Gudmundsson}},\
  }\href@noop {} {\bibfield  {journal} {\bibinfo  {journal} {Journal of
  physics: Condensed matter}\ }\textbf {\bibinfo {volume} {27}},\ \bibinfo
  {pages} {015301} (\bibinfo {year} {2015})}\BibitemShut {NoStop}%
\bibitem [{\citenamefont {Abdullah}\ \emph {et~al.}(2014)\citenamefont
  {Abdullah}, \citenamefont {Tang}, \citenamefont {Manolescu},\ and\
  \citenamefont {Gudmundsson}}]{Nzar116.23}%
  \BibitemOpen
  \bibfield  {author} {\bibinfo {author} {\bibfnamefont {N.~R.}\ \bibnamefont
  {Abdullah}}, \bibinfo {author} {\bibfnamefont {C.-S.}\ \bibnamefont {Tang}},
  \bibinfo {author} {\bibfnamefont {A.}~\bibnamefont {Manolescu}}, \ and\
  \bibinfo {author} {\bibfnamefont {V.}~\bibnamefont {Gudmundsson}},\
  }\href@noop {} {\bibfield  {journal} {\bibinfo  {journal} {Journal of Applied
  Physics}\ }\textbf {\bibinfo {volume} {116}},\ \bibinfo {eid} {233104}
  (\bibinfo {year} {2014})}\BibitemShut {NoStop}%
\bibitem [{\citenamefont {Gudmundsson}\ \emph {et~al.}(2013)\citenamefont
  {Gudmundsson}, \citenamefont {Jonasson}, \citenamefont {Arnold},
  \citenamefont {Tang}, \citenamefont {Goan},\ and\ \citenamefont
  {Manolescu}}]{Vidar61.305}%
  \BibitemOpen
  \bibfield  {author} {\bibinfo {author} {\bibfnamefont {V.}~\bibnamefont
  {Gudmundsson}}, \bibinfo {author} {\bibfnamefont {O.}~\bibnamefont
  {Jonasson}}, \bibinfo {author} {\bibfnamefont {T.}~\bibnamefont {Arnold}},
  \bibinfo {author} {\bibfnamefont {C.-S.}\ \bibnamefont {Tang}}, \bibinfo
  {author} {\bibfnamefont {H.-S.}\ \bibnamefont {Goan}}, \ and\ \bibinfo
  {author} {\bibfnamefont {A.}~\bibnamefont {Manolescu}},\ }\href@noop {}
  {\bibfield  {journal} {\bibinfo  {journal} {Fortschr. Phys.}\ }\textbf
  {\bibinfo {volume} {61}},\ \bibinfo {pages} {305} (\bibinfo {year}
  {2013})}\BibitemShut {NoStop}%
\bibitem [{\citenamefont {Moldoveanu}\ \emph {et~al.}(2010)\citenamefont
  {Moldoveanu}, \citenamefont {Manolescu}, \citenamefont {Tang},\ and\
  \citenamefont {Gudmundsson}}]{PhysRevB.81.155442}%
  \BibitemOpen
  \bibfield  {author} {\bibinfo {author} {\bibfnamefont {V.}~\bibnamefont
  {Moldoveanu}}, \bibinfo {author} {\bibfnamefont {A.}~\bibnamefont
  {Manolescu}}, \bibinfo {author} {\bibfnamefont {C.-S.}\ \bibnamefont {Tang}},
  \ and\ \bibinfo {author} {\bibfnamefont {V.}~\bibnamefont {Gudmundsson}},\
  }\href {\doibase 10.1103/PhysRevB.81.155442} {\bibfield  {journal} {\bibinfo
  {journal} {Phys. Rev. B}\ }\textbf {\bibinfo {volume} {81}},\ \bibinfo
  {pages} {155442} (\bibinfo {year} {2010})}\BibitemShut {NoStop}%
\bibitem [{\citenamefont {Arnold}\ \emph {et~al.}(2015)\citenamefont {Arnold},
  \citenamefont {Tang}, \citenamefont {Manolescu},\ and\ \citenamefont
  {Gudmundsson}}]{2040-8986-17-1-015201}%
  \BibitemOpen
  \bibfield  {author} {\bibinfo {author} {\bibfnamefont {T.}~\bibnamefont
  {Arnold}}, \bibinfo {author} {\bibfnamefont {C.-S.}\ \bibnamefont {Tang}},
  \bibinfo {author} {\bibfnamefont {A.}~\bibnamefont {Manolescu}}, \ and\
  \bibinfo {author} {\bibfnamefont {V.}~\bibnamefont {Gudmundsson}},\ }\href
  {http://stacks.iop.org/2040-8986/17/i=1/a=015201} {\bibfield  {journal}
  {\bibinfo  {journal} {Journal of Optics}\ }\textbf {\bibinfo {volume} {17}},\
  \bibinfo {pages} {015201} (\bibinfo {year} {2015})}\BibitemShut {NoStop}%
\bibitem [{\citenamefont {Jonasson}\ \emph {et~al.}(2012)\citenamefont
  {Jonasson}, \citenamefont {Tang}, \citenamefont {Goan}, \citenamefont
  {Manolescu},\ and\ \citenamefont {Gudmundsson}}]{PhysRevE.86.046701}%
  \BibitemOpen
  \bibfield  {author} {\bibinfo {author} {\bibfnamefont {O.}~\bibnamefont
  {Jonasson}}, \bibinfo {author} {\bibfnamefont {C.-S.}\ \bibnamefont {Tang}},
  \bibinfo {author} {\bibfnamefont {H.-S.}\ \bibnamefont {Goan}}, \bibinfo
  {author} {\bibfnamefont {A.}~\bibnamefont {Manolescu}}, \ and\ \bibinfo
  {author} {\bibfnamefont {V.}~\bibnamefont {Gudmundsson}},\ }\href {\doibase
  10.1103/PhysRevE.86.046701} {\bibfield  {journal} {\bibinfo  {journal} {Phys.
  Rev. E}\ }\textbf {\bibinfo {volume} {86}},\ \bibinfo {pages} {046701}
  (\bibinfo {year} {2012})}\BibitemShut {NoStop}%
\bibitem [{\citenamefont {Abdullah}\ \emph {et~al.}(2013)\citenamefont
  {Abdullah}, \citenamefont {Tang}, \citenamefont {Manolescu},\ and\
  \citenamefont {Gudmundsson}}]{Nzar.25.465302}%
  \BibitemOpen
  \bibfield  {author} {\bibinfo {author} {\bibfnamefont {N.~R.}\ \bibnamefont
  {Abdullah}}, \bibinfo {author} {\bibfnamefont {C.~S.}\ \bibnamefont {Tang}},
  \bibinfo {author} {\bibfnamefont {A.}~\bibnamefont {Manolescu}}, \ and\
  \bibinfo {author} {\bibfnamefont {V.}~\bibnamefont {Gudmundsson}},\
  }\href@noop {} {\bibfield  {journal} {\bibinfo  {journal} {Journal of
  Physics:Condensed Matter}\ }\textbf {\bibinfo {volume} {25}},\ \bibinfo
  {pages} {465302} (\bibinfo {year} {2013})}\BibitemShut {NoStop}%
\bibitem [{\citenamefont {Nakajima}(1958)}]{Nakajima20.948}%
  \BibitemOpen
  \bibfield  {author} {\bibinfo {author} {\bibfnamefont {S.}~\bibnamefont
  {Nakajima}},\ }\href@noop {} {\bibfield  {journal} {\bibinfo  {journal}
  {Prog. of Theor. Phys.}\ }\textbf {\bibinfo {volume} {20}},\ \bibinfo {pages}
  {948} (\bibinfo {year} {1958})}\BibitemShut {NoStop}%
\bibitem [{\citenamefont {Zwanzig}(1960)}]{Zwanzing.33.1338}%
  \BibitemOpen
  \bibfield  {author} {\bibinfo {author} {\bibfnamefont {R.}~\bibnamefont
  {Zwanzig}},\ }\href {\doibase 10.1063/1.1731409} {\bibfield  {journal}
  {\bibinfo  {journal} {The Journal of Chemical Physics}\ }\textbf {\bibinfo
  {volume} {33}},\ \bibinfo {pages} {1338} (\bibinfo {year}
  {1960})}\BibitemShut {NoStop}%
\bibitem [{\citenamefont {Zibold}\ \emph {et~al.}(2007)\citenamefont {Zibold},
  \citenamefont {Vogl},\ and\ \citenamefont {Bertoni}}]{PhysRevB.76.195301}%
  \BibitemOpen
  \bibfield  {author} {\bibinfo {author} {\bibfnamefont {T.}~\bibnamefont
  {Zibold}}, \bibinfo {author} {\bibfnamefont {P.}~\bibnamefont {Vogl}}, \ and\
  \bibinfo {author} {\bibfnamefont {A.}~\bibnamefont {Bertoni}},\ }\href
  {\doibase 10.1103/PhysRevB.76.195301} {\bibfield  {journal} {\bibinfo
  {journal} {Phys. Rev. B}\ }\textbf {\bibinfo {volume} {76}},\ \bibinfo
  {pages} {195301} (\bibinfo {year} {2007})}\BibitemShut {NoStop}%
\bibitem [{\citenamefont {Nielsen}\ and\ \citenamefont
  {Chuang}(2010)}]{Nielsen2010}%
  \BibitemOpen
  \bibfield  {author} {\bibinfo {author} {\bibfnamefont {M.~A.}\ \bibnamefont
  {Nielsen}}\ and\ \bibinfo {author} {\bibfnamefont {I.~L.}\ \bibnamefont
  {Chuang}},\ }\href@noop {} {\emph {\bibinfo {title} {{Quantum computation and
  Quantum Information}}}}\ (\bibinfo {year} {{Cambridge University Press,
  Cambridge, 2010}})\BibitemShut {NoStop}%
\end{thebibliography}
\end{document}